\documentclass[runningheads]{llncs}
\usepackage[T1]{fontenc}
\usepackage{graphicx}

\usepackage{tabularx} 
\usepackage{float}
\usepackage{xcolor}
\usepackage{booktabs}

\begin{document}
\title{Towards Smart Workplaces: Understanding Mood-Influencing Factors of the Physical Workspace in Collaborative Group Settings}
%
\titlerunning{Mood-Influencing Factors in Collaborative Group Settings}

\author{Tzu-Hui Wu\inst{1}\orcidID{0009-0001-4101-0200} \and
Sebastian Cmentowski\inst{1}\orcidID{0000-0003-4555-6187} \and
Yunyin Lou\inst{1}\orcidID{0009-0004-9597-3602} \and
Jun Hu\inst{1}\orcidID{0000-0003-2714-6264} \and
Regina Bernhaupt\inst{1}\orcidID{0000-0003-3522-5990}}
\authorrunning{Wu et al.}

\institute{Eindhoven University of Technology, The Netherlands 
\email{\{t.wu2, s.cmentowski, j.hu, r.bernhaupt\}@tue.nl, y.lou@student.tue.nl}}

\maketitle   
\begin{abstract}
Group mood plays a crucial role in shaping workspace experiences, influencing group dynamics, team performance, and creativity. The perceived group mood depends on many, often subconscious, aspects such as individual emotional states or group life, which make it challenging to maintain a positive atmosphere. Intelligent technology could support mood regulation in physical office environments, for example, as adaptive ambient lighting for mood regulation. However, little is known about the relationship between the physical workspace and group mood dynamics. To address this knowledge gap, we conducted a qualitative user study ($N$=8 workgroups and overall 26 participants) to explore how the physical workspace shapes group mood experiences and investigate employees' perspectives on intelligent mood-aware technologies. Our findings reveal key factors influencing group mood, and participants' expectations for supportive technology to preserve privacy and autonomy. Our work highlights the potential of adaptive and responsive workspaces while also emphasizing the need for human-centered, technology-driven interventions that benefit group well-being.

\keywords{Group Mood  \and Shared Experience \and Group Dynamics \and Physical Workspace \and Intelligent Technology \and Office Well-being.}
\end{abstract}
\section{Introduction}

When a group works with a collective purpose \cite{1}, group mood refers to the shared affective atmosphere experienced by the group members in response to a specific event \cite{2}. This affective sharing process shapes the dynamic, intertwined pattern of emotions within the group \cite{3}. Studies have shown how collective-level affective experiences influence group dynamics in organizational functioning: group effectiveness in decision-making, the attitudes and behaviors of group members, team performance and creativity tasks \cite{4}. Therefore, fostering ``positive workplace vibes'' has become a popular topic in organizational management.

{\setlength{\parskip}{0.5em}
In this context, the physical workspace environment can be seen as a complex system encompassing all objects and stimuli that employees encounter and perceive in their work \cite{35}. Prior work leverages intelligent measurements and interventions to mediate mood experiences and support positive ways of working in office environments. Some tools visualize a user’s arousal level to support a productive vibe \cite{6}. Other researchers designed systems that collect individual data and transform the datasets into collective visualizations to create a reflective process for complete teamwork \cite{7,8}. However, research on understanding the relationship between the group’s experiences and their work environment is limited, with a focus mainly on the individual lens, ignoring the reflection from the collective lens. There is currently no explicit knowledge available to assess the workplace's impact on employees' group experiences. It remains unclear what factors could affect group mood and whether intelligent technology can play a meaningful role in fostering beneficial group moods in work settings.

To address this research gap, we conducted an exploratory study to gain a deep understanding of the potential influence of physical workplace shaping group mood experiences, and participants' expectations and concerns regarding the use of intelligent technology in fostering positive group moods. Specifically, we adopted the user experience curve method and combined it with semi-structured interviews to obtain user responses and envisions based on their lived experiences of complex and multimodal mood experiences. After asking members of 8 workgroups to sketch their meeting mood states, we conducted in-depth interviews to investigate the mood-influencing factors of the workplace and expand on the participants' ideas of how technology can contribute to office well-being. 

The results of our thematic analysis provides an understanding of how the physical workplace influences group mood dynamics, yielding insights that are directly applicable to human-computer interaction (HCI) and workplace design. Specifically, we identified three key aspects at the group level linked to positive mood patterns in office environments: well-structured and low-hierarchy meetings, strong interpersonal connections, and optimal environmental conditions with flexible physical settings. Importantly, these factors are double-edged; if not managed appropriately, they can negatively impact the perception and experience of group mood. These findings offer an experience-oriented perspective grounded in real-world settings. Our analysis also highlights user concerns regarding the integration of intelligent technologies in workspaces, particularly the balance between privacy, autonomy, and system adaptability. In summary, our paper contributes in two ways:
(1) An understanding of relevant aspects in the workplace environment that affect group mood patterns; and
(2) Promising design directions based on participants' expectations and concerns about intelligent technology supporting group mood experiences in physical workspace. }

\section{Related Work}
\subsection{Group Mood}
Group mood can be defined as a pervasive affective atmosphere experienced by group members during a group activity \cite{1}, which emerges within a group, shaped by the interaction and reciprocal influence between individuals \cite{3}. Group affect includes both immediate shared emotional reactions (group emotion) and more enduring shared mood experiences (group mood) \cite{1}. While group emotion refers to direct responses to specific stimuli \cite{11}, group mood encompasses all consistent or uniform emotional responses, representing the general emotional tone \cite{9,10}.

Recently, group mood has been explored within the fields of organizational sciences and human psychology. Prior research focuses mainly on classifying the types and dimensions of mood \cite{1}. Additionally, group mood is generally assessed through self-reported measurements, using the average score to represent the overall group mood \cite{12,13}. However, these studies may not fully capture the collective experience, potentially overlooking the contextual factors, especially the physical workspace, that shape group dynamics. In this research, we focus on group moods in the physical workspace as diffuse and enduring affective states that are shared by group members.

\subsection{HCI and Mood Research}
With the help of HCI approaches, users could receive data-driven self-insights to help optimize their behavioral patterns according to mood patterns \cite{15}. Currently, HCI research uses physiological or behavioral indicators to detect moods. Physiological signals include skin temperature \cite{6,16} and heart rate variability \cite{7,17}, while behavioral signals include facial or verbal expressions \cite{18,19,20}, gestural or bodily movements \cite{21}, and user-product interaction behaviors \cite{22}. Another approach to detecting mood is to use sentiment analysis, inferring moods from texts posted on social media \cite{23}. In addition, several studies enable the detection of mood-related contextual data, such as date and time, location, or weather \cite{24}.

A recurring challenge in the literature is the lack of reliability in mood-related data. Physiological signals, commonly used as proxies for mood, are often unstable due to factors such as sensor invasiveness \cite{25,26} and variable environmental conditions \cite{27}, which undermines their reliability. Additionally, mood self-tracking methods are inherently subjective \cite{25}, and users may inadvertently misrepresent their emotions and experiences \cite{28}. To enhance reliability, some studies have proposed combining automated sensor technology with self-tracking approaches \cite{29,30}. Consequently, researchers have increasingly employed sensor-based systems in physical environments to detect and respond to human emotional states \cite{53}. However, our understanding of how environmental factors relate to mood remains limited. Moreover, concerns about system surveillance and user privacy are widespread, with users often feeling overly monitored by technology \cite{23,31} and expressing apprehension regarding the security and privacy of their mood data \cite{16}. Therefore, it necessitates a deeper understanding of how physical workplace characteristics influence group mood, and for the development of measurement and intervention strategies that enhance mood regulation without compromising user privacy or comfort.

\subsection{Physical Workspace for Office Well-being}
Previous studies explored the potential influence of physical workspace characteristics (e.g., light, sound, air quality) on employees' mental health \cite{35}. The workplace environment is a complex system that encompasses not only objective physical stimuli at work, but also the ways in which individual occupants subjectively perceive these stimuli \cite{44}. For example, research found that exposure to daylight could increase productivity, mood, and reduce fatigue \cite{45}. Increased background noise can also relate to reduced productivity, concentration, and well-being. Although employees prefer to work in an open work environment \cite{46,47}, disturbing background noise could negatively influence peoples' mental health \cite{47,48}. It also seems that products and the spatial features of a room can contribute to certain group moods by affording or limiting interactions. For example, being physically distant from the center of a clustering activity caused some participants to feel ``out of the group vibe'' \cite{1}. A presentation screen also contributed to the group mood by attracting the group’s attention.

\subsection{Intelligent Technology in Physical Workplace for Promoting Office Well-being}
In recent years, many academic disciplines have taken advantage of technologies to promote office well-being. Office well-being covers both psychological and physical aspects, which is considered crucial to work performance, job satisfaction, and employee health \cite{32,33,34}. Intelligent technology has significantly transformed the physical workplace, enabling dynamic interactions that influence office well-being. Enhancing physical activity and reducing mental health problems has become a top priority in the workplace \cite{35,36}. Interactive environments utilize smart devices and responsive systems to facilitate collaboration, adapt to users' needs, and create an active atmosphere, such as visualizing computer-based activity on ambient displays to reduce sedentary behavior \cite{37}, or facilitating walking using interactive screens \cite{38}. Ambient intelligence extends these capabilities by embedding context-aware technologies into the environment, allowing workplaces to sense and respond to users through data collected from internet of things (IoT) devices. For example, personalized ambient lighting systems were used to reduce distraction \cite{39}, and twinkly lights on ambient installation to nudge people to change their behavior \cite{40}. Additionally, digital twins further enhance workplace intelligence by creating virtual replicas of physical spaces \cite{41}. These replicas allow real-time monitoring of the workplace, providing insights into how spatial arrangements, environmental factors (e.g., light, sound, and air quality), and user interactions impact office well-being \cite{42}.

\section{Research Focus}
The goal of our study is to gain a deeper understanding of how the physical workplace shapes group mood dynamics and how intelligent technology could support positive group mood experiences. Prior research highlights the importance of environmental and social factors in influencing office well-being, yet a comprehensive exploration of these influences in real-world office settings remains limited. To design intelligent mood-aware technologies that foster positive group moods, it is essential to first understand users' perceptions on how the physical workplace influences their group mood. Accordingly, our research question is:

\vspace{0.5em}

\noindent \textbf{RQ1: How does the physical workspace influence group mood?}

\vspace{0.5em}

\noindent To answer this question, we subdivide it into three sub-questions that focus on the general factors, the influence of meeting stages, and the impact on the group mood:

\begin{itemize}
    \item \textbf{RQ1a:} Which factors of the physical workspace influence group mood?
    \item \textbf{RQ1b:} Which factors are important during which stage of group work?
    \item \textbf{RQ1c:} Which factors are facilitators or inhibitors for group mood?
\end{itemize}

\noindent Additionally, previous research has extensively explored both group mood definitions and intelligent workplace technologies as separate domains. However, the intersection of these two areas---how intelligent technology can actively shape and support group mood in office environments---remains underexplored. Existing studies on workplace technology primarily focus on individual well-being and productivity, often overlooking the collective emotional experience of workgroups. Therefore, our research investigates employees' expectations and concerns towards the integration of intelligent technology in fostering positive group moods. This leads us to our second research question:

\vspace{0.5em}

\noindent \textbf{RQ2: How do users perceive the potential of Mood-Aware Workspace Environment (MAWEs) for enhancing mood and well-being in group scenarios?} 

\vspace{0.5em}

\noindent Our exploration is a crucial first step toward designing intelligent interventions to enhance office well-being. By identifying the mood-influencing factors and understanding participants' expectations towards intelligent technologies, we aim to develop actionable guidelines for creating mood-aware workspace environments that foster a positive atmosphere and promote office well-being.
                      
\section{Method}
\begin{figure}
\includegraphics[width=\textwidth]{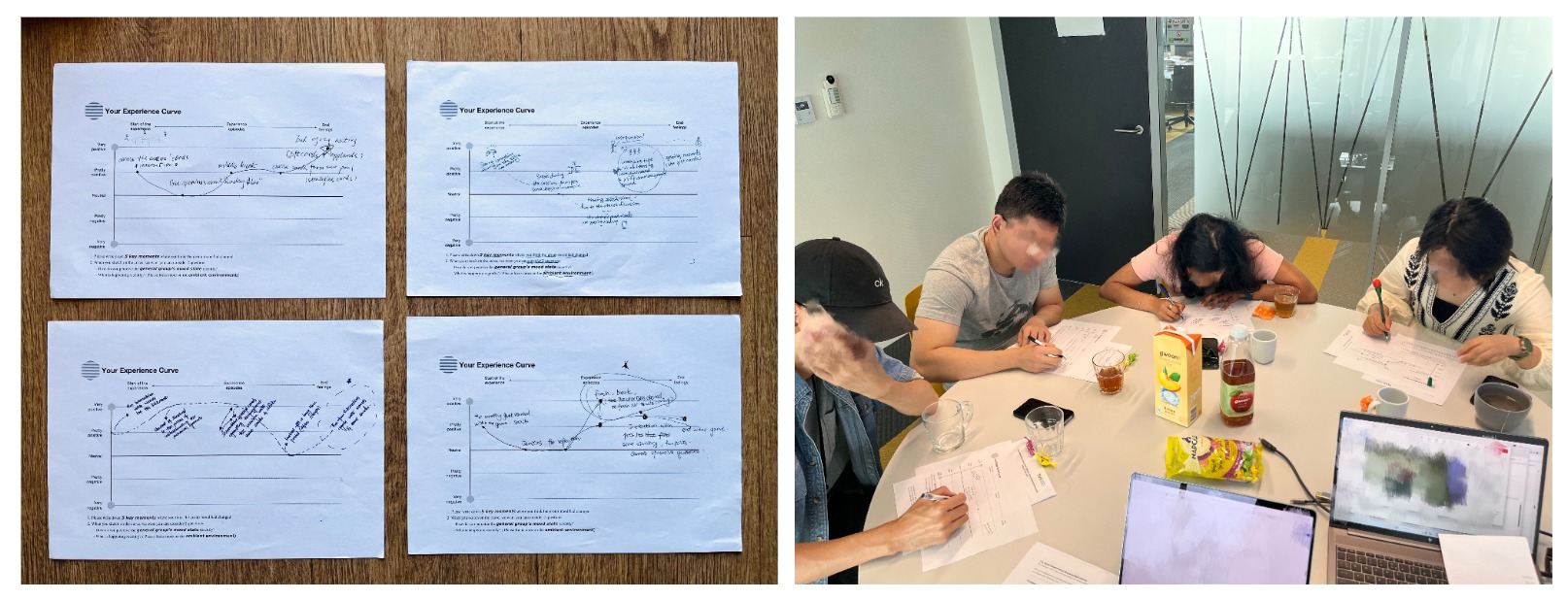}
\caption{Research materials used for self-reporting: The experience curve (left) and study setting (right).} \label{fig1}
\end{figure}

For our user study, we combined two methodologies, the user experience curve (UX Curve) and semi-structured interviews, to answer our research questions (see Fig.~\ref{fig1}). The UX Curve  is ``a method to support users in retrospectively reporting how their experience changed over time'' \cite{49}. It helps capture subjective and self-reported experiences with a product. The goal is to gather reports of the experience as it unfolds, thereby supporting memory reconstruction. In this study, group members self-reported their experienced group mood development based on their actual meeting. In this context, the UX curve can reveal subtle mood changes over the course of an experience, providing rich contextual insights. Using these sketches as a starting point, we then conducted in-depth interviews to explore the factors influencing group mood patterns and delve deeper into participants’ perspectives on how technology could enhance their office well-being. The combination of UX curves and semi-structured interviews captures both immediate mood experiences and deeper insights behind these experiences.

\subsection{Participant}
For our study, we recruited 8 groups (a total of 26 participants) engaged in ongoing group projects from the University. Our sample size aligns with prior qualitative research norms by Braun and Clarke \cite{55,braun2021one}, who indicate that thematic saturation is typically reached with 12–25 participants. Three of the groups consisted of PhD staff who met regularly for academic research. The other groups were ongoing student design teams working on group assignments in workshops or focus groups. To accurately capture group mood experiences, we established sampling criteria based on factors that influence mood convergence in workgroups \cite{9,50}: Membership stability, task interdependence, and social interdependence. Additionally, when groups exceed six members, individuals are less likely to engage in collective interactions that foster mood homogeneity \cite{1,51}. Based on these criteria, we selected groups with a maximum of 6 members, who engaged in an ongoing project (membership stability), performed distinct tasks while collaborating within the group (task interdependence), and had established familiarity with one another (social interdependence). Our final set comprised 8 groups with 3-4 participants each, who engaged in 60-90 minutes of group work. 

We recognize the importance of personal and contextual factors in influencing group mood. To control for external influences, we collected detailed contextual information, including meeting type, physical environment, and collaboration duration. Table~\ref{tab1} gives an overview of the demographic and contextual information of participants.

\begin{table}[ht]
\centering
\caption{Participant group profiles: group number, size, participants (coded), meeting types, environment, collaboration, participant roles in each session.}\label{tab1}
  \setlength{\tabcolsep}{6pt}

\resizebox{\textwidth}{!}{
  \begin{tabular}{c c c c c c}
  \toprule
  {\bfseries Number} & {\bfseries Size}& {\bfseries Meeting type} & {\bfseries Environment} & {\bfseries Duration}& {\bfseries Participant Roles}\\
  \midrule
  Group 1 & 3 & Focus group & Meeting room & < 3 months & 1 facilitator, 2 members\\
  Group 2 & 4 & Student project & Open space & One month & 4 members\\
  Group 3 & 3 & Academic meeting & Meeting room & > 6 months & 2 supervisors, 1 PhD\\
  Group 4 & 3 & Academic meeting & Meeting room &  > 3 months & 2 supervisors, 1 PhD\\
  Group 5 & 3 & Academic meeting & Meeting room & > 6 months & 1 supervisor, 2 PhD\\
  Group 6 & 4 & Focus group & Meeting room  &  > 6 months & 1 facilitator, 3 members\\
  Group 7 & 3 & Workshop & Open space & > 6 months & 1 facilitator, 2 members\\
  Group 8 & 3 & Focus group & Meeting room & < 3 months & 1 facilitator, 2 members\\
  \bottomrule
  \end{tabular}
  }
\vspace{1em}
\end{table}

\subsection{Research Procedure and Materials}
\begin{figure}
\includegraphics[width=\textwidth]{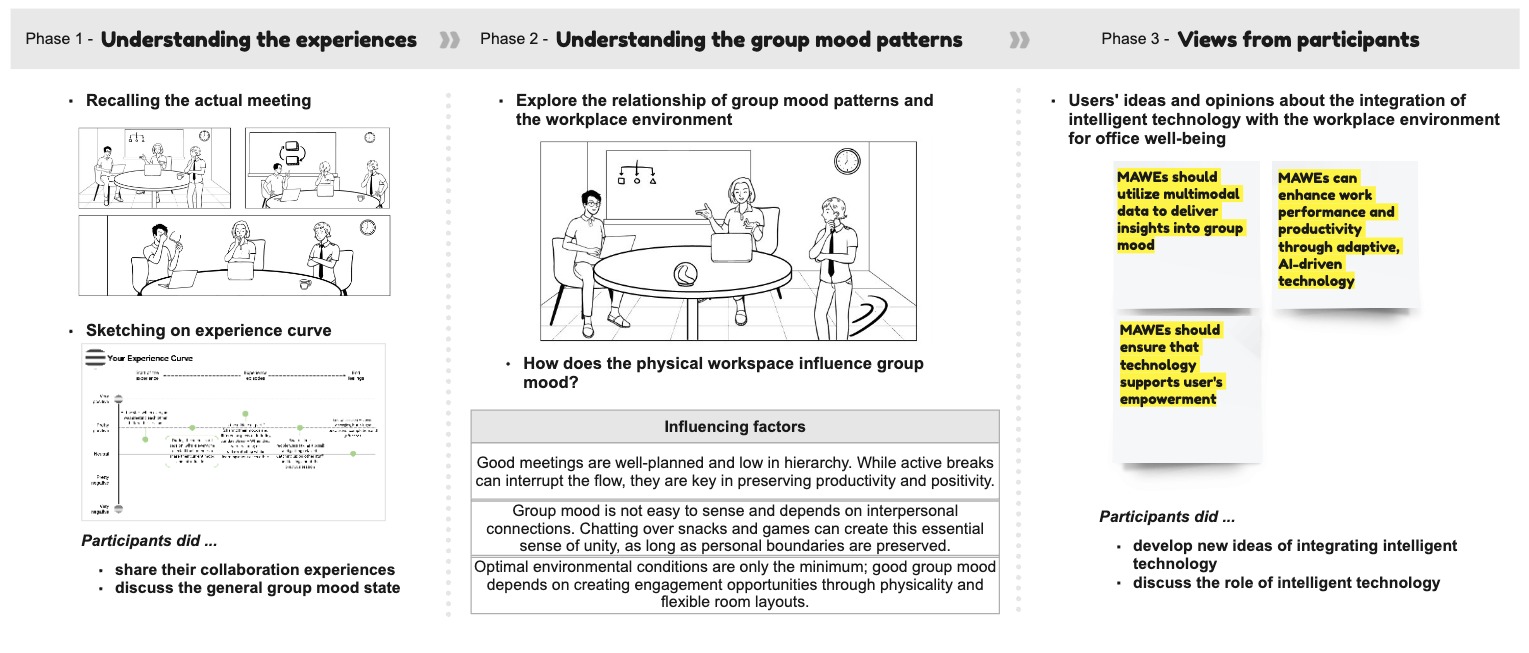}
\caption{A visual outline of the study, including provided materials, tasks for participants, and transcribed data in each phase.} \label{fig2}
\end{figure}

The procedure of this study is described in Fig.~\ref{fig2}. 
Following each group meeting, the 3-4 participants attended a shared interview session with the first author. We deliberately chose group interviews as they encourage participants to build on each other’s ideas, revealing insights that might not emerge in isolation. While one-on-one sessions allow for deep personal reflections, they might fail to capture the ideas that emerge from dynamic collaborative discussion. These sessions lasted 60-70 minutes and consisted of two parts: First, participants recalled their meeting by sketching and describing their perception of the group mood development using a pen-and-paper experience curve. Afterwards, the researcher conducted a semi-structured interview session to explore commonalities and distinctions between group members’ mood perceptions and their perspectives on intelligent mood-aware technologies.
In detail, our interview session used the following structure:

\subsubsection{Part 1: Retrospecting the actual meetings and sketching the experience curve.} This first part aimed to understand the actual experiences of group collaboration. We introduced the experience curve to the participants and asked them to share their collaboration experiences by sketching the perceived group mood development and identifying five key moments where they perceived a significant change in group mood. Data collection in this phase consisted of their sketches and descriptions.

\subsubsection{Part 2: Semi-structured interview.} The second part focused on understanding the group mood patterns and exploring participants' views on using intelligent technologies to enhance group office well-being. The first part aimed to explore the factors shaping group mood dynamics in the workplace. We asked guiding questions about their experience curves, such as ``What do you think caused the change in group mood at this key moment?'', ``Do you observe any connection between this key moment and the surrounding context?'', and ``What do you see as a main trigger in the physical workplace that affects your group mood?''. We also explored participants' differing perceptions of the experience curves, asking follow-up questions to understand the factors influencing their varied interpretations of group mood. For example, we inquired, ``Why did you choose a different group mood state at this moment?''. For the second part of the interview, we introduced the concept of a ``Mood-Aware Workspace Environment'' (MAWEs) and invited participants to share their perspectives. We asked open-ended questions such as ``What are your thoughts on integrating intelligent technologies into the workplace to create a mood-aware environment? How might these technologies impact your experience?''.

Naturally, qualitative data like ours is specific to the interviewed participants and might be influenced by personal circumstances and external events. We took particular care to minimize these influences by framing the interview questions explicitly on mood-related factors of the physical workplace, distinguishing them from external events, and asking participants to focus primarily on environmental factors. When external influences were reported, we separated these cases in our analysis, ensuring that only mood changes attributed to workspace conditions were captured.
All participants provided with informed consent prior to the study. Upon the consent of the participants, all the sessions were voice recorded.

\subsection{Data Analysis}

For data analysis, the main author first transcribed the audio recordings using Microsoft Teams\footnote{\url{https://www.microsoft.com/en-us/microsoft-teams/group-chat-software}}, then reviewed and cleaned the original transcripts before initiating the analysis. 
For the subsequent coding of both the interview data and the experience curve sketches, we used the online platform Dovetail\footnote{\url{https://dovetail.com/}}. Therefore, we followed a six-step thematic analysis process originally outlined by Braun and Clarke \cite{52}. While we consider our analysis procedure generally as ``reflexive'', marked by iterations in which the codes evolve and change during the analysis~\cite{braun2021one}, we also incorporated aspects of other strains of thematic analysis: two researchers coded all data independently (i.e., coding reliability) and iteratively reviewed and aligned the results codes frequently (i.e., codebook approach). To start the process, we first derived an initial codebook deductively, drawing from group-mood-influencing factors reported in prior literature~\cite{1}. Additionally, we tagged the identified factors according to their occurrence in the meeting (e.g., beginning, middle, end) and their impact on the experience curve (i.e., facilitating or inhibiting group mood). For the perceptions of mood-aware systems, we solely relied on inductive codes. 

Two researchers independently coded 25\% of the interview data before meeting with a third, more senior author to discuss and unify the resulting codebook. Afterwards, we repeated the same procedure for the next 25\% of the data, further refining the codebook. For the remaining data, the two main coders followed this procedure without the involvement of the third author. At each discussion stage, we discussed all coded data, resolved any coding differences, and revised/consolidated the codebook. Please refer to Appendix for a picture of the final codebook. After finalizing coding, the first author used an affinity mapping approach to group the codes and generate initial themes. Subsequently, the three primarily involved authors discussed and refined the results to establish the final theme definitions. The analysis aimed to achieve two objectives: (1) identifying factors that shape group mood dynamics in the workspace, and (2) investigating participants' expectations regarding intelligent  technology that support group experiences in workspace settings.

\section{Results}
In this section, we present the six main themes we identified within the data through thematic analysis (see Table~\ref{tab2}). In the following parts, we will elaborate each theme and participant quotes. Participants are labeled ``P'' followed by group number, for example, ``P3-2'' for the second participant in the third group.

\begin{table}
\caption{Table of the six main themes extracted from the data alongside their related subthemes, separated by research question.}\label{tab2}
\centering
\setlength{\tabcolsep}{2pt}
\renewcommand{\arraystretch}{1.2}
\begin{tabularx}{\textwidth}{X X} 
\toprule
{\bfseries Themes} & {\bfseries Subthemes} \\
\midrule
(1) Good meetings are well-planned and low in hierarchy. While active breaks can interrupt the flow, they are key in preserving productivity and positivity
& {\itshape 1.1. The meeting type and organization determines the importance and direction of group mood developments;\newline
1.2. Breaks, Surprises, and Physical Activity are important but can be distracting }\\
\hline
(2) Group mood is not easy to sense and depends on interpersonal connections. Chatting over snacks and games can create this essential sense of unity, as long as personal boundaries are preserved
& {\itshape 2.1. Perceiving group mood is hard but communication and social bonds lead to mood convergence eventually;\newline
2.2. Communication, informal chats, and group interactions are essential for creating a productive environment} \\
\hline
(3) Optimal environmental conditions are only the minimum; positive group mood depends on creating engagement opportunities through physicality and flexible room layouts.
& {\itshape 3.1. Calm, well-lit, and ventilated rooms are basic requirements for productive work;\newline
3.2. Working with physical materials and communication tools creates an engaging atmosphere;\newline
3.3. Room layouts are a two-edged sword: adaptable layouts benefit while unfamiliarity hinders engagement }\\
\hline
(4) MAWEs should utilize multimodal data to deliver insights into group mood, ensuring context-aware communication of mood to enhance understanding and cohesion
& {\itshape4.1. MAWEs should leverage multimodal data to gain comprehensive insights into group mood dynamics;\newline
4.2. MAWEs should support context-aware and subtle communication of emotional sharing }\\
\hline
(5) MAWEs can enhance work performance and productivity through adaptive, AI-driven technology while ensuring explainable and transparent communication to understand system's behaviors
& {\itshape 5.1. Adaptive interventions in MAWEs boost energy and re-engage individuals, thereby enhancing work performance;\newline
5.2. AI meeting tools in MAWEs boost productivity with effective time tracking and management }\newline
5.3. MAWEs should offer flexible, explainable, and transparent information to ensure understanding in system decisions\\
\hline
(6) MAWEs should ensure that technology supports user empowerment in environmental control without compromising ethical and privacy concerns
&{\itshape 6.1. MAWEs should empower users to control environmental settings, balancing automated regulation with group consensus;\newline
6.2. Ensuring clear social boundaries and carefully managing sensitive data are crucial for addressing ethical and privacy concerns in MAWEs }\\
\bottomrule
\end{tabularx}
\end{table}


\subsection{Theme 1: Good meetings are well-planned and low in hierarchy. While active breaks can interrupt the flow, they are key in preserving productivity and positivity.}

\subsubsection{Subtheme 1.1: The meeting type and organization determines the importance and direction of group mood developments.} Most participants emphasized that the structure and type of meetings significantly determine how group mood evolves: \textit{``Perception of group mood would change depending on whether the meeting is more casual or formal''} (\textbf{P5-2}). Participants also reported that well-organized meetings with a clear agenda and low hierarchy not only promote inclusivity, but also shape the emotional tone of the group. For example, participants' experiences and perceptions of the group mood and dynamics could vary depending on their specific role (e.g., moderator/instructor). In contrast, group mood suffers from low energy levels due to long meetings with excessive workload and poor scheduling: \textit{``Busted for a long time period''} \textbf{(P2-4)} and \textit{``Getting tired due to the intense discussion''} \textbf{(P2-1)}. 

\subsubsection{Subtheme 1.2: Breaks, Surprises, and Physical Activity are important but can be distracting.} While many participants recognized that breaks, surprises, and physical activity are critical to recharging and energizing the group, they also mentioned that these elements can sometimes be distracting if not well-timed. Participants noted that incorporating physical activity during meetings is key to maintaining a positive mood and gauging individual engagement: \textit{``It's positive because we were moving around rather than just staying in the same place''} \textbf{(P1-1)}, highlighting how even small bursts of body movement can enhance the overall atmosphere. Similarly, breaks, though sometimes reluctant, were seen as necessary for recharging energy: \textit{``When we come back from the break, we felt a bit more relaxed''} (\textbf{P8-3}). However, novelty and relief during meetings is a double-edged sword. While participants appreciated surprises to break the routine, several noted that unanticipated challenges could disturb the workflow: \textit{``Our workshop was interrupted because other people also want to use the table''} \textbf{(P7-1)}.

\subsection{Theme 2: Group mood is not easy to sense and depends on interpersonal connections. Chatting over snacks and games can create this essential sense of unity, as long as personal boundaries are preserved.}

\subsubsection{Subtheme 2.1: Perceiving group mood is hard but communication and social bonds lead to mood convergence eventually.} Participants expressed that discerning individual and group mood can be challenging, as it is a complex combination of individual moods and how they affect the group dynamics: \textit{``It was really hard to differentiate my own mood from the group mood''} \textbf{(P4-3)}. However, they also highlighted that effective communication---positive feedback, supportive tone, timely rewards, and successful outcomes---can significantly enhance the perception of group mood: \textit{``Our moods were influenced by the tone of the instructor's speech and outcomes''} \textbf{(P6-2)}. Furthermore, strong team bonds, shared backgrounds, and compatible personalities were also identified as critical factors that minimize information gaps. Through consistent communication and the strengthening of social bonds, individual moods eventually align: \textit{``Our mood gradually aligned to a similar level in group meeting''} \textbf{(P3-2)}. This convergence highlights the gradual but powerful role of interpersonal communication in shaping a shared emotional atmosphere. 

\subsubsection{Subtheme 2.2: Communication, informal chats, and group interactions are essential for creating a productive environment.} Some participants pointed out that informal interactions are essential for fostering unity. They noted that active social engagement and regular group interactions with lightweight, positive topics tend to elevate group mood: \textit{``The vibe was pretty positive, ending with the fun interaction and discussion with memes card'' }\textbf{(P2-4)}. While sensitive or intrusive conversations related to \textit{``more personal things''} \textbf{(P1-2)} can lead to discomfort. Participants also highlighted shared interests and mutual expectations as key to creating a sense of unity, thereby enhancing communication: \textit{``The important thing is that we have a really clear goal, [...] so everything just smoothly.''} \textbf{(P2-3)}. This shows that while group mood is hard to capture, it can be nurtured through intentional, informal communication and strong social bonds, which in turn contribute to a more productive work environment.
Providing snacks and games fosters a relaxed atmosphere, encouraging group members to connect on a more personal level

\subsection{Theme 3: Optimal environmental conditions are only the minimum; good group mood depends on creating engagement opportunities through physicality and flexible room layouts.}

\subsubsection{Subtheme 3.1: Calm, well-lit, and ventilated environments are basic requirements for productive work.} Participants stressed that optimal environmental conditions---such as thermal comfort, optimal lighting, and fresh air---play a crucial role in promoting relaxation and elevating group mood. One participant explained that \textit{``The air is becoming not good at all, [...] staying here in a hot environment, which is not helpful for attention and mood''} \textbf{(P3-2)}, illustrating the direct impact of environmental conditions on concentration. Additionally, while ambient sounds and background music can stimulate creativity, excessive noise was described as a major distraction undermining group mood and productivity: \textit{``It's ok if there just regular ambient sound, but I get distracted if someone is screaming'' }\textbf{(P7-1)}. These conditions, although fundamental, are viewed as minimum requirements for a productive work environment.

\subsubsection{Subtheme 3.2: Working with physical materials and communication tools creates an engaging atmosphere.} Beyond the basic conditions, participants highlighted the use of physical materials, such as cards and sticky notes, to be engaging and helpful in facilitating the group discussion and problem solving process. They felt that the tangible nature of these materials allowed them to better organize their thoughts and collaborate effectively: \textit{``fun interaction with memes cards and stickers'' }\textbf{(P2-4)}. Likewise, the use of shared communication tools, like collaborative whiteboards and digital displays, was also noted as a key factor in fostering active participation: \textit{``When we start to open or use some cooperative tools, like using the Miro or whiteboard, [...] that's quite a positive atmosphere''} \textbf{(P5-2)}. Integrating both physical materials and collaborative tools can sustain group engagement and creativity.

\subsubsection{Subtheme 3.3: Room layouts are a double-edged sword: adaptable layouts benefit while unfamiliarity hinders engagement.} Most participants discussed the role of room layouts, revealing a double-edged effect. On the one hand, flexible configuration and adaptable layouts benefit engagement, communication, and concentration. For example, a U-shaped or circular table arrangement \textbf{(P4-3)} promotes interaction, as it encourages participants to face each other and engage in more discussions. On the other hand, unfamiliar environments and physical distance impact the individual perception of the group mood. Getting used to the environment \textbf{(P5-3)} and collaborating to address issues contributed to an improved group dynamic over time. Conversely, physical distant affects the perception: \textit{``I feel distancing, [...] feel a little bit bad''} \textbf{(P4-1)}. Maintaining a sense of familiarity is important to ensure a positive atmosphere.

\subsection{Theme 4: MAWEs should utilize multimodal data to deliver insights into group mood, ensuring context-aware communication of mood to enhance understanding and cohesion.}

\subsubsection{Subtheme 4.1. MAWEs should leverage multimodal data to gain comprehensive insights into group mood dynamics.} Participants proposed some examples of using various sensors and data points, such as group distribution and positioning, physiological indicators like heart rate and skin conductance, sound, body status and behaviors, to infer group atmosphere. They believe that these physical cues can provide insights into the overall group dynamics. Some participants also suggested the possibility of allowing people to self-report their emotional states: \textit{``Self-rating emotions could help, [...] offer some insights''} \textbf{(P5-1)}. But there are concerns about the reliability of self-reporting, as people may not always accurately perceive or be willing to share their true emotional state. In addition, some participants highlighted the role of embodied physical objects in capturing interactive mood-related data. For example, \textit{``talking objects''} \textbf{(P8-2)} and \textit{``perceptible plants''} \textbf{(P8-3)} could display a general group mood indicator without being too invasive.

\subsubsection{Subtheme 4.2. MAWEs should support context-aware and subtle communication of emotional sharing.} Participants wished for systems that are be able to identify when individuals are inclined to share personal information and are willing to reveal their feelings. They expressed a desire to know others' moods, as it could help them gauge if they have done something wrong or if the other person is upset: \textit{``I want to know people's feelings when I'm suspicious or I do something wrong''} \textbf{(P2-3)}. However, there are also concerns about the sensitivity of sharing emotions, especially if it could be recognized by supervisors or used for performance evaluations: \textit{``Mood should not always be visible. When you had difficult conversation with your supervisor, I'm not sure if you want them to know that you're uncomfortable in the moment''} \textbf{(P4-3)}. The overall sentiment is that while knowing others' moods could be beneficial in some situations, people should still have control over whether they want to share their emotions. Some participants explore ideas like using plants or other representations to provide subtle cues about the group mood or energy levels. These insights indicate that context-aware emotional communication not only preserve personal boundaries but also enhances team cohesion by facilitating a balance between individual privacy and group mood awareness.

\subsection{Theme 5: MAWEs can enhance work performance and productivity through adaptive, AI-driven technology while ensuring explainable and transparent communication to understand system’s behaviors.}

\subsubsection{Subtheme 5.1. Adaptive interventions in MAWEs boost energy and re-engage individuals, thereby enhancing work performance.} Participants envisioned adaptive intelligent interventions that could dynamically adjust to their needs, reducing friction in collaborative work. As articulated by \textbf{P4-2}, physical settings adjusting automatically can boost energy levels. Some noted MAWEs should support adaptive workspaces: \textit{``The chairs or tables in the room can automatically be configured depending on what we need''} \textbf{(P4-2)}. Some participants also noted that adaptive interventions, including physical activity suggestions, mood-tailored music, and the display of mood changes, help re-engage distracted individuals. However, concerns about automation were raised, with participants preferring that systems act as ``assistants'' rather than imposing mandatory interventions.

\subsubsection{Subtheme 5.2. AI meeting tools in MAWEs boost productivity with effective time tracking and management.} Participants want AI-powered features that can manage real-time meeting processes, such as summarizing discussions or searching for information, to enhance meeting productivity. \textbf{P3-2} envisioned a virtual facilitator that summarizes key points and suggests discussion topics for the group, helping to maintain a productive meeting atmosphere. They also emphasize the importance of having reminder systems \textbf{(P7-1)} to help keep meetings on track and ensure engagement, especially in larger group settings where it can be difficult to interrupt someone who is taking too long. 

\subsubsection{Subtheme 5.3. MAWEs should offer flexible, explainable, and transparent information to ensure understanding in system decisions.}~Most participants highlighted the desire for physical workspaces that provide both flexibility and explainable information to support understanding. The potential for MAWEs, such as automated curtains, to act unpredictably or annoying users is a concern, especially when their actions lack clear explanations: \textit{``It might be nice if it could indicate in some way why the curtains are closing or opening''}~\textbf{(P4-2)}. Participants also noted need for flexibility and transparency in the work environment, such as information on how each person's state affects the overall group dynamics. Group level is a complex social phenomenon, with certain individuals having a stronger influence on the group mood than others.

\subsection{Theme 6: MAWEs should ensure that technology supports user's empowerment in environmental control without compromising ethical and privacy concerns.}

\subsubsection{Subtheme 6.1. MAWEs should empower users to control environmental settings, balancing automated regulation with group consensus.} Some participants pointed out the challenges of MAWEs that need to be addressed. Users expressed frustration with the automated curtain and lighting system in their office. They found the curtains to be either too bright or too dark, and the automatic adjustments to be disruptive, especially during meetings. There is a lack of group control over the curtains, leading to conflicts between colleagues who have different preferences for lighting. Users suggest that a more customizable or zoned lighting system could help address these issues and better accommodate the diverse needs of office workers \textbf{(P1-2).} The concept of user control in group settings raises the question of whether intelligent technologies can enhance group mood dynamics without undermining individual autonomy. One participant explained that \textit{“The environment could understand the general mood in the office, but it shouldn’t force interactions or mood changes”}~\textbf{(P6-2)}.

\subsubsection{Subtheme 6.2. Ensuring clear social boundaries and carefully managing sensitive data is crucial for addressing ethical and privacy concerns in MAWEs.} Most participants were concerned about privacy issues relating to sharing personal information, such as heart rate and mood, with a smart environment. They preferred that the environment provides only abstract or general information about mood levels, rather than specific personal details, ensuring the social boundaries. Intelligent technologies should provide sufficient flexibility and options to allow teams to decide when and how interventions or information occur. \textbf{P6-1} noted, \textit{``I don’t want to let environment knows my personal information like heart rate''}. Participants were also concerned about the sensitivity of personal information, especially in a work context where it could be accessed by leader and potentially influence perceptions or decisions.

\section{Discussion}

The results in Section 5 summarize our findings regarding the relevant aspects in the workplace environment that shape the group mood patterns, as well as participants' expectation and concerns about MAWEs. In this section we discuss the interpretation and application of these results. 

\subsection{RQ1: How does the physical workspace influence group mood?}
\textbf{Theme 1} reveals that the type of meeting and its organizational structure fundamentally shape the group's emotional tone. This finding is supported by existing literature: the meeting type, group leadership and attendee roles may determine the expected group mood \cite{1,9}. However, our study extends these insights by highlighting the role of physical activities and breaks in sustaining productivity and positivity. According to our participants, the structure of a meeting, combined with observable physical activities, provides valuable cues about individual and collective status, enabling designers to better infer the overall group atmosphere. Breaks and surprise elements serve as facilitator, energizing participants and reinforcing a positive group vibe. But if breaks and unexpected challenges are not well-timed, they could be inhibitors of group mood. \textbf{Theme 2} also identifies that perceiving group mood is difficult, yet strong interpersonal interactions and social bonds foster a productive environment that gradually leads to mood convergence. This finding is in line with prior work by Chung and Grèzes \cite{3}, who demonstrate how individual emotions, through continuous emotional interactions, align into a unified group mood. Answering \textbf{RQ1c}, we identified that effective group interactions and chatting over snacks and games can serve as facilitators for building familiarity and team cohesion, not only boosting group mood but also creating a strong sense of unity. At the same time, personal boundaries can lower group mood, as sensitive or intrusive discussions can lead to discomfort. \textbf{Theme 3} emphasizes the importance of creating engagement through physicality and flexible room layouts while maintaining optimal environmental conditions. This finding resonate with prior research: physical workplace characteristics significantly impact employees' mental health (e.g. stress, fatigue, or mood) \cite{35}. But our results extend the literature by revealing that adaptive, interactive workspaces act as a double-edged sword: when well-managed, they facilitate positive group dynamics, but an unfamiliar workspace can hinder productivity. With regards to \textbf{RQ1a}, we expand our understanding of factors influencing group mood by revealing that specific physical workplace elements, well-structured and low-hierarchy meetings, strong interpersonal connections, and optimal environmental conditions paired with flexible physical settings, play a crucial role in shaping group mood. 

Furthermore, our analysis reveals that the significance of these factors fluctuates over time (\textbf{RQ1b}). The initial stage of group work is especially sensitive to the meeting type and the clarity of its structure \textbf{(Theme 1)}, with a sense of novelty providing an additional energy boost. Participants noted that the meeting types and initial tasks establish the foundational emotional tone for the group. Interestingly, many described the default state at the beginning as a neutral mood---neither distinctly positive nor negative---serving as a baseline from which emotions and engagement can later shift up or down depending on the group dynamics and activities. Conversely, during the middle and later stages of group work, the physical environment and collaborative tools take on an essential role in fostering engagement and interactivity. Participants indicated that optimal environmental conditions, the use of physical materials, and well-designed room layouts create an atmosphere that not only fosters interaction but also re-energizes the group \textbf{(Theme 3)}. Additionally, effective communication, robust group interactions, and strong social bonds become more vital for sustaining energy in the later stages \textbf{(Theme 2)}. However, these factors are not isolated. They are intricately intertwined, collectively shaping the dynamic fluctuations of group mood and lead to mood alignment eventually.

\subsection{RQ2: How do users perceive the potential of MAWEs for enhancing mood and well-being in group scenarios?}
Participants have two expectations for MAWEs: utilizing multimodal data to infer and communicate group mood, discussed in \textbf{Theme 4}, and leveraging AI-driven technology to enhance work performance, discussed in \textbf{Theme 5}. These expectations demonstrate promising examples of using various sensors to infer and interpret group mood. Our discussion also highlights potential challenges and limitations of these approaches, including the difficulty in accurately interpreting emotional states, the risk of making incorrect assumptions, and the necessity of balancing privacy concerns with the goal of fostering a responsive and supportive work environment. Embodied physical objects that combine sensors and self-reporting methods can provide a more precise assessment of group mood. Theme 5 also emphasized adaptive and AI-driven technology to improve office well-being and productivity in a shared work setting by monitoring and responding to their emotional and energy states. Sometimes we passively accept the system's information and interventions, overlooking the autonomy individuals should have in the physical workspace. In such cases, MAWEs look like a black box, hiding the reasons behind their actions and leaving users in the dark about system's behaviors. This opacity not only undermines trust, but also hinders understanding of the technology. Participants also raised challenges of MAWEs: supporting user empowerment in environmental control without compromising ethical and privacy concerns, discussed in \textbf{Theme 6}. Privacy, transparency, and user autonomy remain key concerns. Although users are generally open to smart, adaptive workspaces, their effectiveness depends on a balance between automated regulation, user empowerment, and robust ethical data management. Achieving this balance is essential for building trust and ensuring technology supports, rather than undermine, human interaction and agency. 

\subsection{Design Implications}
Finally, we propose actionable recommendations for designing effective MAWEs based on our findings.

\textbf{Enhancing Group Mood Monitoring with Reliability and Privacy Control.} Our findings suggest that designers can combine a diverse array of data, including physiological signals, spatial positioning, audio cues, self-reports, body language, posture, and interactions with furniture or devices, to accurately capture group mood \textbf{(Subtheme 4.1)}. Previous work highlights that mood detection via sensors or self-tracking is often unreliable, as physiological and behavioral signals can be unstable across contexts and users \cite{27}. By integrating this heterogeneous data using advanced analytics, MAWEs can identify subtle emotional shifts and trigger adaptive interventions when energy levels drop or stress increases. Prior research has used ``SocialStools'' to promote group cohesion through embodied interaction by measuring the distance, orientation and interaction among group members \cite{54}. However, designers should address privacy concerns: ``always-on'' mood tracking could lead to perceived privacy invasion \textbf{(Subtheme 6.2)}. To protect user comfort and agency, we propose the anonymous sharing of physiological data and the use of non-intrusive, ambient sensors or indirect mood indicators (e.g., behavior patterns or embodied physical objects), which can infer group mood without collecting sensitive personal data.

\textbf{Fostering Mood Alignment.} Our qualitative findings suggest facilitating the alignment of individual moods into a group emotional state through intentional design interventions. Mood alignment, a specific kind of group dynamics, cultivates a sense of togetherness and strengthens collaborative efforts \cite{3,54}. Possible interventions could aim for more mood awareness in collaborative settings, which can trigger reflection \cite{23}, lead to behavioral mimicry, and ultimately foster mood alignment among team members \cite{57}. Additionally, designers could define appropriate timing and prompt cues to invite participants to share their feelings at predetermined moments \textbf{(Subtheme 4.2)}. By guiding when and how group express themselves, whether via physical tools or digital notifications, these measures help team members adjust their behaviors in response to others, fostering a stronger sense of unity and enhancing collaboration.

\textbf{Empowering Group Control in Adaptive Workspaces.} The feedback from our participants also suggests empowering groups through shared, adaptive control over their workspace environments, ensuring interventions align with collective preferences and individual comfort \textbf{(Subthemes 5.1, 6.1)}. Mood-regulating recommendations should be context-aware and sensitive to individual differences, allowing groups to modify interventions according to their preferences. A collaborative control interface, such as a group dashboard or mobile app, that presents information about ambient conditions (e.g., light level, temperature, noise) allows team members to suggest or vote on adjustments, which fosters transparency and group consensus, reducing the feeling of control by automated systems. In addition, although automated adaptations can be effective in regulating group mood, designers should consider system intrusiveness and unintended disruptions to workflow or mood. Previous research has shown that unobtrusive ambient displays can effectively regulate affect without distracting users \cite{39}. To address these concerns, interventions should aim for seamless integration into existing work routines or unobtrusive ambient mood indicators, minimizing distraction and avoiding unnecessary interactions.

\subsection{Future Work and Limitation}
While our findings offer valuable insights into how the physical workspace influences group mood, this qualitative study has limitations. Primarily, qualitative methods may not fully capture the variability in group experiences across diverse real-life settings. Our participant pool ($N$=26 across 8 groups) was relatively homogeneous, potentially limiting the generalizability of our conclusions. Given that no new codes or themes emerged during analysis of the final two interviews, we are confident saturation was reached; however, larger and more varied samples could offer additional knowledge.
Additionally, our analysis could reflect interpretive biases given our diverse academic backgrounds: two authors were from industrial design and the third from computer science with experience in VR gaming research. While these backgrounds potentially introduced biases, they also provided a comprehensive understanding of group mood dynamics.
Future research should incorporate larger, diverse participant groups to enhance generalizability and use controlled experiments to empirically validate specific environmental and interpersonal factors identified in this study. Addressing privacy concerns, we recommend developing privacy-preserving behavioral analysis methods to infer group mood, using anonymized data on posture, spatial positioning, and interaction patterns. Validation studies involving controlled experiments and longitudinal field studies in realistic workplace environments are also recommended to evaluate the effectiveness and long-term impacts of workspace interventions on group mood dynamics.

\section{Conclusion}
Group mood in the workspace is inherently difficult to describe, making it equally challenging to measure or design for. While prior research has largely focused on individual experiences, the impact of physical workspaces on collaborative group affect remains underexplored. Our study addresses this gap by uncovering how workspace features shape group dynamics. Before we can design interventions that can enhance group mood, we must first understand the underlying factors and expectations. Our qualitative findings provide this foundational knowledge. Through an exploratory study employing the experience curve and semi-structured interviews with eight groups, our analysis revealed that a variety of factors, i.e., well-structured and low-hierarchy meetings, strong interpersonal connections, and optimal environmental conditions paired with flexible physical settings, significantly influence group mood. Our findings further emphasized the temporal differences of mood fluctuations, highlighting how these factors can either facilitate or inhibit group dynamics. These factors should not be seen in isolation, but as interconnected forces shaping the emergence of group mood. Subsequently, we explored the potential of intelligent technologies to support mood-aware workspaces, suggesting that effective MAWEs must balance adaptive, automated interventions with strong user autonomy and transparent communication. Based on participants' feedback, we propose a set of design implications aimed at guiding future efforts in multimodal data assessment, mood alignment, and the empowerment of autonomy. In the context of HCI research, our study contributes a novel perspective on group well-being in shared environments. It offers practical value for researchers and designers through actionable insights and design recommendations, particularly around how smart technologies might support communication and adapt dynamically to group mood.




%
%


\begin{credits}
\subsubsection{\ackname} The first author is partially supported by the Guangzhou Elites Scholarship Council. 

\end{credits}
%
%
%
\bibliographystyle{splncs04}
\bibliography{Reference}
%




\end{document}